\documentclass{llncs}
\usepackage[english]{babel}
\usepackage[utf8x]{inputenc}
\usepackage[T1]{fontenc}
\usepackage{amsfonts}
\usepackage{amssymb}
\usepackage{mathtools}
\usepackage{listings}

\usepackage{amsmath}
\usepackage{graphicx}
\usepackage{solarized-light}
\usepackage{xcolor}
\usepackage[colorinlistoftodos]{todonotes}
\usepackage[colorlinks=true, allcolors=blue]{hyperref}
\usepackage{cleveref}
\usepackage{inconsolata}
\usepackage{euler}
\usepackage{xspace}
\usepackage{enumitem}
\usepackage{microtype}
\usepackage{subfiles}

\definecolor{amaranth}{rgb}{0.9, 0.17, 0.31}
\title{Ephemeral Data Handling in Microservices}
\subtitle{Technical Report}

\author{
Saverio Giallorenzo\inst{1} \and
Fabrizio Montesi\inst{1} \and
Larisa Safina\inst{1,2} \and 
Stefano Pio Zingaro\inst{3}
}

\institute{
  $^1$University of Southern Denmark,
  $^2$Innopolis University,
  $^3$Università di Bologna/INRIA
}

\clearpage{}%

\let\emptyset\varnothing
\newcommand{\framework}{\textsf{TQuery}\xspace}
\newcommand{\gram}{::=}
\newcommand{\Div}{\;\;|\;\;}
\newcommand{\trueHl}[1]{\colorbox{gray!15}{$#1$}}
\newcommand{\hl}[1]{#1}
\newcommand{\emptyseq}{\epsilon}
\newcommand{\emptytree}{\tau}
\newcommand{\emptyarray}{\alpha}
\newcommand{\emptyval}{\mathbb\upsilon}

\newcommand{\Array}[2]{\left[\, #1,\cdots,#2\,\right]}

\newcommand{\size}{\texttt{\#}}

\newcommand{\tint}{\texttt{int}}
\newcommand{\tstr}{\texttt{str}}
\newcommand{\tsort}{\texttt{sort}}

\newcommand{\pt}[1]{#1\ . \ }
\newcommand{\ept}[1]{#1}
\newcommand{\apt}[2]{[\hspace{-.15em}[\, {#2}\, ]\hspace{-.15em}]^{#1}}
\newcommand{\eval}[2]{#1\!\!\downarrow\!\!#2}

\newcommand{\multOp}{\mathbf{E}}
\newcommand{\mult}[3]{\multOp(#1,#2)^{#3}}

\newcommand{\app}[1]{\triangleright#1}

\newcommand{\func}[1]{\textbf{\texttt{#1}}}
\newcommand{\evalDef}{\func{eval}}
\newcommand{\merge}{\oplus}

\makeatletter
\DeclareRobustCommand\myop[1]{%
  \mathop{\vphantom{\sum}\mathpalette\myop@{#1}}\slimits@
}
\newcommand{\myop@}[2]{%
  \vcenter{%
    \sbox\z@{$#1\sum$}%
    \hbox{\resizebox{\ifx#1\displaystyle.8\fi\dimexpr\ht\z@+\dp\z@}{!}
{$\m@th#2$}}%
  }%
}
\makeatother

\newcommand{\bigop}[1]{\DOTSB\myop{#1}}

\newcommand{\match}{\mu}
\newcommand{\sat}{\models}
\newcommand{\project}{\pi}
\newcommand{\unwind}{\omega}
\newcommand{\group}{\gamma}
\newcommand{\lookup}{\lambda}

\newcommand{\true}{\mathtt{true}}
\newcommand{\false}{\mathtt{false}}

\newcommand{\lstmath}[1]{\mbox{\lstinline{#1}}}

\lstdefinelanguage{jolie}{
    basicstyle=\normalfont\ttfamily,
    numbers=left,
    stepnumber=1,
    numbersep=8pt,
    frame=none,
    morekeywords=
    {_,any,||,undef,match,unwind,project,group,lookup,is_defined,exists},
    string=[s]{"}{"},
    keywords = [2]{in,if,by,else},
    keywordstyle=[2]{\color{sorange}\bfseries},
    comment=[l]{//},
    literate=
     *{0}{{{\color{smagenta}0}}}{1}
      {1}{{{\color{smagenta}1}}}{1}
      {2}{{{\color{smagenta}2}}}{1}
      {3}{{{\color{smagenta}3}}}{1}
      {4}{{{\color{smagenta}4}}}{1}
      {5}{{{\color{smagenta}5}}}{1}
      {6}{{{\color{smagenta}6}}}{1}
      {7}{{{\color{smagenta}7}}}{1}
      {8}{{{\color{smagenta}8}}}{1}
      {9}{{{\color{smagenta}9}}}{1}
      {\{}{{{\bfseries\color{sorange}{\{}}}}{1}
      {\}}{{{\bfseries\color{sorange}{\}}}}}{1}
      {[}{{{\bfseries\color{sblue}{[}}}}{1}
      {]}{{{\bfseries\color{sblue}{]}}}}{1}
      {:}{{{\bfseries\color{sorange}{:}}}}{1}
      {,}{{{\bfseries\color{sorange}{,}}}}{1}
      {+}{{{\bfseries\color{sorange}{+}}}}{1}
      {*}{{{\bfseries\color{sorange}{*}}}}{1}
      {-}{{{\bfseries\color{sorange}{-}}}}{1}
      {!}{{{\bfseries\color{sorange}{!}}}}{1}
      {||}{{{\bfseries\color{sorange}{||}}}}{1}
      {<}{{{\bfseries\color{sorange}{<}}}}{1}
      {>}{{{\bfseries\color{sorange}{>}}}}{1}
      {==}{{{\bfseries\color{sorange}{==}}}}{2}
      {\\}{{{\bfseries\color{sorange}{\textbackslash}}}}{1}
      {\&\&}{{{\bfseries\color{sorange}{\&\&}}}}{1}
      {@}{{{\bfseries\color{sorange}{@}}}}{1}
      {;}{{{\bfseries\color{sorange}{;}}}}{1}
}

\newcommand\scalemath[2]{\scalebox{#1}{\mbox{\ensuremath{\displaystyle #2}}}}
\newcommand\mathexample[1]{\noindent\colorbox{sbase3}{\begin{minipage}
{\textwidth}#1\end{minipage}}}
\newcommand\castToSet[1]{\left\{\!\left\{#1\right\}\!\right\}}

\newcommand{\chainop}{\textbf{\texttt{\color{sorange}|>}}\xspace}

\clearpage{}%

\begin{document}
\maketitle

\begin{abstract} In modern application areas for software systems --- like
eHealth, the Internet-of-Things, and Edge Computing --- data is encoded in
heterogeneous, tree-shaped data-formats, it must be processed in real-time,
and it must be ephemeral, i.e., not persist in the system. While it is
preferable to use a query language to express complex data-handling logics,
their typical execution engine, a database external from the main application,
is unfit in scenarios of ephemeral data-handling. A better option is
represented by integrated query frameworks, which benefit from existing
development support tools (e.g., syntax and type checkers) and execute within
the application memory. In this paper, we propose one such framework that, for
the first time, targets tree-shaped, document-oriented queries. We formalise
an instantiation of MQuery, a sound variant of the widely-used MongoDB query
language, which we implemented in the Jolie language. Jolie programs are
microservices, the building blocks of modern software systems. Moreover, since
Jolie supports native tree data-structures and automatic management of
heterogeneous data-encodings, we can provide a uniform way to use MQuery on
any data-format supported by the language. We present a non-trivial use case
from eHealth, use it to concretely evaluate our model, and to illustrate our
formalism.
\end{abstract}

\section{Introduction}
Modern application areas for software systems---like eHealth~\cite{orszag2008evidence}, the
Internet of Things~\cite{Baker2017}, and Edge Computing~\cite{Shi2016}---need to address
two requirements: velocity and variety~\cite{mehta2004handbook}. Velocity
concerns managing high throughput and real-time processing of data. Variety
means that data might be represented in heterogeneous formats, complicating
their aggregation, query, and storage.
Recently, in addition to velocity and variety, it has become increasingly
important to consider \emph{ephemeral} data handling
\cite{tene2012big,shein2013ephemeral}, where data must be processed in
real-time but not persist --- ephemeral data handling can be seen as the
opposite of dark data~\cite{darkdata}, which is data stored but not used. The
rise of ephemeral data is due to scenarios with heavy resource constraints
(e.g., storage, battery) --- as in the Internet of Things and Edge Computing
--- or new regulations that may limit what data can be persisted, like the
GDPR~\cite{Mostert2015} --- as in eHealth.

Programming data handling correctly can be time consuming and error-prone with
a general-purpose language. Thus, often developers use a query language, paired
with an engine to execute them~\cite{cheney2013practical}. When choosing the
query execution engine, developers can either \emph{A}) use a database
management system (DBMS) executed outside of the application, or \emph{B})
include a library that executes queries using the application memory.

Approach \emph{A}) is the most common. Since the early days of the Web,
programmers integrated application languages with relational (SQL-based) DBMSs
for data persistence and manipulation~\cite{welling2003php}. This pattern
continues nowadays, where relational databases share the scene with new
NoSQL~\cite{mehta2004handbook} DBMSs, like MongoDB~\cite{mongodb} and Apache
CouchDB~\cite{couchdb}, which are document-oriented. Document-oriented
databases natively support tree-like nested data structures (typically in the
JSON format). Since data in modern applications is typically structured as
trees (e.g., JSON, XML), this removes the need for error-prone
encoding/decoding procedures with table-based structures, as in relational
databases. However, when considering ephemeral data handling, the issues of
approach \emph{A}) overcome its benefits even if we consider NoSQL DBMSs:

\begin{enumerate}[label={\textcircled{\texttt{\footnotesize\arabic*}}}]

    \item{\label{prob:drivers}\emph{Drivers and Maintenance}.} An external DBMS is an
    additional standalone component that 
    needs to be installed, deployed, and maintained. To interact with the DBMS,
    the developer needs to import in the application specific drivers
    (libraries, RESTful outlets). As with any software
    dependency, this exposes the
    applications to issues of version incompatibility \cite{dependency_hell}.

    \item{\label{prob:security}\emph{Security Issues}.} The companion DBMS is
    subject to weak security configurations~\cite{mongodb_security} and query
    injections, increasing the attack surface of the application.
    
    \item{\label{prob:tools}\emph{Lack of Tool Support}.}
    Queries to the external DBMS are typically black-box entities (e.g.,
    encoded as plain strings), making them opaque to analysis tools
    available for the application language (e.g., type checkers)
    \cite{cheney2013practical}.

    \item{\label{prob:velocity}\emph{Decreased Velocity and Unnecessary
    Persistence}.} Integration bottlenecks and overheads degrade the velocity
    of the system. Bottlenecks derive from resource constraints and slow
    application-DB interactions; e.g., typical database connection
    pools \cite{visveswaran2000dive} represent a potential bottleneck in the
    context of high data-throughput.
    Also, data must be inserted in the database and eventually
    deleted to ensure ephemeral data handling. Overheads also come in the form
    of data format conversions (see item~\ref{prob:variety}).

    \item{\label{prob:variety}\emph{Burden of Variety}.} The DBMS typically
    requires a specific data format for communication, forcing the programmer
    to develop ad-hoc data transformations to encode/decode data in transit (to
    insert incoming data and returning/forwarding the result of queries).
    Implementing these procedures is cumbersome and error-prone.

\end{enumerate}

On the other side, approach \emph{B}) (query engines running within the application)
is less well explored, mainly because of the historical bond between query languages and persistent data storage. However, it holds potential for ephemeral data handling.
Approach \emph{B}) avoids issues~\ref{prob:drivers} and~\ref{prob:security} by
design. Issue~\ref{prob:tools} is sensibly reduced, since both queries and data
can be made part of the application language. Issue~\ref{prob:velocity} is also
tackled by design. There are less resource-dependent bottlenecks and no
overhead due to data insertions (there is no DB to populate) or deletions (the
data disappears from the system when the process handling it terminates). Data
transformation between different formats (item~\ref{prob:variety}) is still an
issue here since, due to variety, the developer must convert incoming/outgoing
data into/from the data format supported by the query engine. Examples of
implementations of approach \emph{B}) are
LINQ~\cite{meijer2006linq,cheney2013practical} and CQEngine~\cite{cqengine}.
While LINQ and CQEngine grant good performance (velocity), variety is still an
issue. Those proposals either assume an SQL-like query language or rely on a
table-like format, which entail continuous, error-prone conversions between
their underlying data model and the heterogeneous formats of the
incoming/outgoing data.

\paragraph{Contribution.}
Inspired by approach \emph{B}), we implemented a framework for ephemeral data
handling in microservices; the building blocks of software for our application
areas of interest. Our framework includes a query language and an execution
engine, to integrate document-oriented queries into the
Jolie~\cite{MGZ14,jolieweb} programming language. The language and our
implemented framework are open-source projects\footnote{
\url{https://github.com/jolie/tquery}}. Our choice on Jolie comes
from the fact that Jolie programs are natively microservices~\cite{DGLMMMS17}.
Moreover, Jolie has been successfully used to build
Internet-of-Things~\cite{GGLZ18} and eHealth~\cite{wareflo} architectures, as
well as Process-Aware Information Systems~\cite{montesi2016process}, which
makes our work directly applicable to our areas of interest. Finally, Jolie
comes with a runtime environment that automatically translates
incoming/outgoing data (XML, JSON, etc.) into the native, tree-shaped data
values of the language --- Jolie values for variables are always trees. By
using Jolie, developers do not need to handle data conversion themselves,
since it is efficiently managed by the language runtime. Essentially, by being
integrated in Jolie, our framework addresses issue~\ref{prob:variety} by
supporting \emph{variety by construction}.

As main contribution of this paper, in \Cref{sec:formalisation}, we present
the formal model, called \framework, that we developed to guide the
implementation of our Jolie framework. \framework is inspired by
MQuery~\cite{botoeva18}, a sound variant of the MongoDB Aggregation
Framework~\cite{agg_framework}; the most popular query language for NoSQL data
handling. The reason behind our formal model is twofold. On the one hand, we
abstract away implementation details and reason on the overall semantics of
our model --- we favoured this top-down approach (from theory to practice) to
avoid inconsistent/counter-intuitive query behaviours, which are instead
present in the MongoDB Aggregation Framework (see~\cite{botoeva18} for
details). On the other hand, the formalisation is a general reference for
implementors; this motivated the balance we kept in \framework between formal
minimality and technical implementation details --- e.g., while
MQuery adopts a set semantics, we use a tree semantics.

As a second contribution, in \Cref{sec:use_case} we present a non-trivial
eHealth use case to overview the \framework operators, by means of their Jolie
programming interfaces. The use case is also the first concrete evaluation of
MQuery and, in \Cref{sec:formalisation}, we adopt the use case as our running
example to illustrate the semantics of \framework.

\section{A Use Case from eHealth}
\label{sec:use_case}

In this section, we illustrate our proposal with an eHealth use case taken
from~\cite{Vigevano2018}, where the authors delineate a diagnostic algorithm
to detect cases of encephalopathy. The handling follows the principle of
``data never leave the hospital'' in compliance with the GDPR~
\cite{ROSE20141212}. In the remainder of the paper, we use the use case to
illustrate the formal semantics of \framework. Hence, we do not show here the
output of \framework operators, which are reported in their relative
subsections in \cref{sec:formalisation}.
While the algorithm described in~\cite{Vigevano2018} considers a plethora of
clinical tests to signal the presence of the neurological condition, we focus
on two early markers for encephalopathy: fever in the last 72 hours and
lethargy in the last 48 hours. That data is collectible by
commercially-available smart-watches and smart-phones~\cite{bunn2018current}:
body temperature and sleep quality. We report in \Cref{lst:usecase_data}, in a
JSON-like format, code snippets exemplifying the two kinds of data structures.
At lines 1--2, we have a snippet of the biometric data collected from the
smart-watch of the patient. At lines 4--6 we show a snippet of the sleep
logs~\cite{Thurman2018}. Both structures are arrays, marked \lstinline{[ ]},
containing tree-like elements, marked \lstinline|{ }|. At lines 1--2, for each
\lstinline{date} we have an array of detected temperatures (\lstinline{t}) and
heart-rates (\lstinline{hr}). At lines 4--6, to each year (\lstinline{y})
corresponds an array of monthly (\lstinline{M}) measures, to a month
(\lstinline{m}), an array of daily (\lstinline{D}) logs, and to a day
(\lstinline{d}), an array of logs (\lstinline{L}), each representing a sleep
session with its start (\lstinline{s}), end (\lstinline{e}) and quality
(\lstinline{q}).

\begin{figure}[tb]
\begin{lstlisting}[language=jolie,basicstyle=\ttfamily\footnotesize,%
caption=Snippets of biometric (line 1) and sleep logs (lines
3--5) data.,label=lst:usecase_data]
[{date:20181129,t:[37,...],hr:[64,...]},
 {date:20181130,t:[36,...],hr:[66,...]},...]

[{y:2018,M:[...,{m:11,D:[{d:29,L:[{s:"21:01",e:"22:12",q:"good"},
{s:"22:36",e:"22:58",q:"good"},...]},{d:30,L:[
{s:"20:33",e:"22:12",q:"poor"},...]},...]},...]},...]
\end{lstlisting}
\vspace{-2em}
\end{figure}

On the data structures above, we define a Jolie microservice, reported in
\Cref{lst:usecase}, which describes the handling of the data and the workflow
of the diagnostic algorithm, using our implementation of
\framework. The example is detailed enough to let us illustrate all the
operators in \framework: \lstinline{match}, \lstinline{unwind}, 
\lstinline{project}, \lstinline{group}, and \lstinline{lookup}. Note that,
while in \Cref{lst:usecase} we hard-code some data (e.g., integers
representing dates like \lstinline{20181128}) for presentation purposes, we
would normally use parametrised variables.

In \Cref{lst:usecase}, line 1 defines a request to an external service,
provided by the \textbf{\texttt{HospitalIT}} infrastructure. The service
offers functionality \lstinline{getPatientPseudoID} which, given some
identifying \lstinline{patientData} (acquired earlier), provides a
pseudo-anonymised identifier --- needed to treat sensitive health data ---
saved in variable \lstinline{pseudoID}. 

At lines 2--6 (and later at lines 9--17) we use the chaining operator
$\chainop$ to define a sequence of calls, either to external services, marked
by the \lstinline{@} operator, or to the internal \framework library. The
$\chainop$ operator takes the result of the execution of the expression at its
left and passes it as the input of the expression on the right.

At lines 2--6 we use \framework operators \lstinline{match} and
\lstinline{project} to extract the recorded temperatures of the
patient in the last 3 days/72 hours.

At line 2 we evaluate the content of variable \lstinline{credentials}, which
holds the certificates to let the Hospital IT services access the
physiological sensors of a given patient.
In the program, \lstinline{credentials} is passed by the chaining
operator at line 3 as the input of the external call to functionality
\lstinline{getMotionAndTemperature}. That service call returns the biometric
data (\Cref{lst:usecase_data}, lines 1--2) from the
\textbf{\texttt{SmartWatch}} of the patient. While the default syntax of
service call in Jolie is the one with the double pair of parenthesis (e.g., at
line 1 \Cref{lst:usecase}), thanks to the chaining operator $\chainop$ we can
omit to specify the input of \lstinline{getMotionAndTemperature} (passed by
the $\chainop$ at line 3) and its output (the biometric data exemplified at
\Cref{lst:usecase_data}) passed to the $\chainop$ at line 4. At line 4 we use
the \framework operator \lstinline{match} to filter all the entries of the
biometric data, keeping only those collected in the last 72 hours/3 days
(i.e., since \lstinline{20181130}). The result of the \lstinline{match} is
then passed to the \lstinline{project} operator at line 5, which removes all
nodes but the temperatures, found under \lstinline{t} and renamed
\lstinline{in} \lstinline{temperatures} (this is required by the interface of
functionality \lstinline{detectFever}, explained below). The
\lstinline{project}ion also includes in its result the \lstinline{pseudoID} of
the patient, \lstinline{in} node \lstinline{patient_id}. We finally store
(line 6) the prepared data in variable \lstinline{temps} (since it will be
used both at line 7 and 16).

At line 7, we call the external functionality \lstinline{detectFever} to
analyse the temperatures and check if the patient manifested any fever,
storing the result in variable \lstinline{hasFever}.

\begin{figure}[t]
\begin{lstlisting}[basicstyle=\ttfamily\footnotesize,caption=Encephalopathy
Diagnostic Algorithm.,label=lst:usecase]
getPatientPseudoID@$\textbf{HospitalIT}$( patientData )( pseudoID );
credentials 
$\chainop$ getMotionAndTemperature@$\textbf{SmartWatch}$
$\chainop$ match { date == 20181128 || date == 20181129 || date == 20181130 } 
$\chainop$ project { t in temperatures, pseudoID in patient_id }
$\chainop$ temps;
detectFever@$\textbf{HospitalIT}$( temps )( hasFever );
if( hasFever ){
  credentials
  $\chainop$ getSleepPatterns@$\textbf{SmartPhone}$
  $\chainop$ unwind  { M.D.L }
  $\chainop$ project{y in year,M.m in month,M.D.d in day,M.D.L.q in quality}
  $\chainop$ match { year == 2018 && month == 11 && ( day == 29 || day == 30 ) }
  $\chainop$ group   { quality by day, month, year }
  $\chainop$ project { quality, pseudoID in patient_id }
  $\chainop$ lookup  { patient_id == temps.patient_id in temps }
  $\chainop$ detectEncephalopathy@$\textbf{HospitalIT}$   }
\end{lstlisting}
\vspace{-2em}
\end{figure}

After the analysis on the temperatures, \lstinline{if} the patient
\lstinline{hasFever} (line 8), we continue
testing for lethargy. To do that, at
lines 9--10, we follow the same strategy described for lines 2--3 to pass the
\lstinline{credentials} to functionality \lstinline{getSleepPatterns}, used to
collect the sleep logs of the patient from her \texttt{\textbf{SmartPhone}}.
Since the sleep logs are nested under years, months, and days, to filter the
logs relative to the last 48 hours/2 days, we first flatten the structure
through the \lstinline{unwind} operator applied on nodes \lstinline{M.D.L}
(line 11). For each nested node, separated by the dot (\lstinline{.}), the
\lstinline{unwind} generates a new data structure for each element in the
array reached by that node. Concretely, the array returned by the
\lstinline{unwind} operator at line 11 contains all the sleep logs in the
shape:
{\small\mathexample{\lstmath{[}\qquad\lstmath{\{year:2018, M:[\{m:11, D:[
\{d:29,L:[\{s:"21:01",e:"22:12",q:"good"\}]\}]\}]\},}}
\hspace{1em}\mathexample{\qquad\;\;\lstmath{\{year:2018, M:[\{m:11, D:[
\{d:29,L:[\{s:"22:36",e:"22:58",q:"good"\}]\}]\}]\}}\quad\lstmath{]}}}
where there are as many elements as there are sleep logs and the arrays under
\lstinline{M}, \lstinline{D}, and \lstinline{L} contain
only one sleep log. Once flattened, at line 12 we modify the data-structure
with the \lstinline{project} operator to simplify the subsequent chained
commands: we rename the node \lstinline{y} \lstinline{in}
\lstinline{year}, we move and rename the node \lstinline{M.m} \lstinline{in}
\lstinline{month} (bringing it at the same nesting level of \lstinline{year});
similarly, we move \lstinline{M.D.d}, renaming it \lstinline{day}, and we move
\lstinline{M.D.L.q} (the log the quality of the sleep), renaming it 
\lstinline{quality} --- \lstinline{M.D.L.s} and
\lstinline{M.D.L.e}, not included in the \lstinline{project}, are discarded.
On the obtained structure, we filter the sleep logs relative to the last 48
hours with the \lstinline{match} operator at line 13. At line 14 we use the
\lstinline{group} operator to aggregate the \lstinline{quality} of the sleep
sessions recorded in the same day (i.e., grouping them
\lstinline{by} \lstinline{day}, \lstinline{month}, and
\lstinline{year}). Finally, at line 15 we select, through a 
\lstinline{project}ion, only the aggregated values of \lstinline{quality} 
(getting rid of \lstinline{day}, \lstinline{month}, and \lstinline{year}) and
we include under node \lstinline{patient_id} the \lstinline{pseudoID} of the
patient. That value is used at line 16 to join, with the \lstinline{lookup}
operator, the obtained sleep logs with the previous values of temperatures
(\lstinline{temps}). The resulting, merged data-structure is finally passed to
the \textbf{\texttt{HospitalIT}} services by calling the functionality
\lstinline{detectEncephalopathy}. 
\section{\framework Framework}
\label{sec:formalisation}

In this section, we define the formal syntax and semantics of the operators of
\framework. We begin by defining data trees:
\[
T \ni t \gram b \ \{ k_i : a_i \}_i
\qquad\qquad
A \ni a \gram \Array{t_1}{t_n}
\]
Above, each tree $t \in T$ has two elements. First, a \emph{root} value $b$,
$b \in \tsort$, where $\tsort = \tstr \cup \tint \cup \cdots \cup
\{\emptyval\}$ and $\emptyval$ is the null value. Second, a set of
one-dimensional vectors, or arrays, containing sub-trees. Each array is
identified by a label $k \in K$. We write arrays $a \in A$ using the standard
notation $\Array {t_1}{t_n}$. We write $k(t)$ to indicate the extraction of
the array pointed by label $k$ in $t$: if $k$ is present in $t$ we return the
related array, otherwise we return the null array $\emptyarray$, formally
\[
k(\ b\ \{k_i:a_i\}_i\ ) = \begin{cases}
a & \mbox{if } (k:a) \in \{k_i:a_i\}_i
\\
\emptyarray & \mbox{otherwise}
\end{cases}
\]
We assume the range of arrays to run from the minimum index $1$ to the maximum
$\size a$, which we also use to represent the size of the array. We use the
standard index notation $a[i]$ to indicate the extraction of the tree at index
$i$ in array $a$. If $a$ contains an element at index $i$ we return it,
otherwise we return the null tree $\emptytree$.
\[
a[i] = \begin{cases}
t_i & \mbox{if } a = \Array{t_1}{t_n} \ \wedge\ 1 \leq i \leq n
\\
\emptytree & \mbox{otherwise}
\end{cases}
\]
\begin{example}[Data Structures]\label{ex:data_structure}
To exemplify our notion of trees, we model the data structures in
\Cref{lst:usecase_data}.
\begin{lstlisting}[basicstyle=\ttfamily\footnotesize]
[ $\emptyval$ {date:[20181129 {}],t:[37{},...],hr:[64{},...]},
  $\emptyval$ {date:[20181130 {}],t:[36{},...],hr:[66{},...]},...]

[$\emptyval${y:[2018{}],M:[$\emptyval${m:[11{}],D:[
  $\emptyval${d:[29{}],L:[$\emptyval${s:["21:01"{}],e:["22:12"{}],q:["good"{}]},...]},
  $\emptyval${d:[30{}],L:[$\emptyval${s:["20:33"{}],e:["22:12"{}],q:["poor"{}]},...]},
...]},...]},...]
 \end{lstlisting}
Note that tree roots hold the values in the data structure (e.g., the
integer representation of the date \lstinline{20181128}). When root values are
absent, we use the null value $\emptyval$.
\end{example}
We define paths $p \in P$ to express tree traversal:
$
P \ni p \gram \pt{e}\ept{p} \Div \emptyseq
$.
Paths are concatenations of expressions $e$, each assumed to evaluate to a
tree-label, and the sequence termination $\emptyseq$ (often omitted in
examples). The application of a path $p$ to a tree $t$,
written $\apt{p}{t}$ returns an array that contains the sub-trees reached
traversing $t$ following $p$. This is aligned with the behaviour
of path application in MQuery which return a set of trees.
In the reminder of the paper, we write $\eval{e}{k}$ to indicate
that the evaluation of expression $e$ in a path results into the label $k$.
Also, both here and in MQuery paths neglect array indexes: for a given path
$e.p$, such that $\eval{e}{k}$, we apply the subpath $p$ to all trees pointed
by $k$ in $t$.
We use the standard array concatenation operator $::$ where $[t_1,\cdots,t_n]=
[t_1]::\cdots::[t_n]$.
We can finally define $\apt{t}{p}$, which either returns an array of trees or
the null array $\emptyarray$ in case the path is not applicable.
\[
\begin{array}{lll}
\apt{t}{p} = \begin{cases}
\apt{t_1}{p'}\ ::\ \cdots\ ::\ \apt{t_n}{p'}
& \mbox{if }
  p = e.p' \ \wedge \
  \eval{e}{k} \ \wedge \ k(t) = [t_1,\cdots,t_n]
  \\
  [\ t\ ] & \mbox{if } p = \emptyseq
  \\
  \emptyarray & \mbox{otherwise}
  \end{cases}
  \end{array}
  \]
In the
reminder, we also assume the following structural equivalences
 \[
\emptyarray :: \emptyarray \equiv \emptyarray
\;\;\quad
\emptyarray :: [\,] \equiv [\,] :: \emptyarray \equiv [\,]::[\,] \equiv [\,]
\;\;\quad
\emptyarray :: a \equiv a :: \emptyarray \equiv [\,] :: a \equiv a :: [\,]
\equiv a
\]

\begin{example}
\label{example:path}
Let us see some examples of path-tree application where we assume a tree $t =
$
\lstinline|$\emptyval$ { x: [ $\emptyval$ { z: [ 1 {}, 2 {} ] , y: [3 {}] } ] }| 
\begin{lstlisting}
$\apt{t}{x.\emptyseq}\ \Rightarrow$ [$\emptyval${z:[1{},2{}],y:[3{}]}]
$\apt{t}{x.z.\emptyseq}\ \Rightarrow$ [1{},2{}]
\end{lstlisting}
\end{example}
We first present the syntax of \framework and then dedicate a subsection to
the semantics of each operator and to the running examples that illustrate
its behaviour.

As in MQuery, a \framework query is a sequence of stages $s$ applied on an
array $a$: $\trueHl{a\app{s}\cdots\app{s}}$. The staging operator $\app{}$ in
\framework is similar to the Jolie chaining operator $\chainop$: they evaluate
the expression on their left, passing its result as input to the expression at
their right. We report in \Cref{fig:syntax} the syntax of \framework,
which counts five stages. The \emph{match} operator $\trueHl{\match_\varphi}$
selects trees according to the criterion $\varphi$. Such criterion is either
the boolean truth $\true$, a condition expressing the equality of the
application of path $p$ and the array $a$, a condition expressing the equality
of the application of path $p_1$ and the application of a second path $p_2$,
the existence of a path $\exists p$, and the standard logic connectives
negation $\neg$, conjunction $\wedge$, and disjunction $\vee$.
The \emph{unwind} operator $\trueHl{\unwind_p}$ flattens an array reached
through a path $p$ and outputs a tree for each element of the array.
The \emph{project} operator $\trueHl{\project_{\Pi}}$ modifies trees by
projecting away paths, renaming paths, or introducing new paths, as described
in the sequence of elements in $\Pi$, which are either a path $p$ or a value
definition $d$ inserted into a path $p$. Value definitions are either: a
boolean value $b$ ($\true$ or $\false$), the application of a path $p$, an 
array of value definitions, a criterion $\varphi$ or the ternary expression,
which, depending the satisfiability of criterion $\varphi$ selects either
value definition $d_1$ or $d_2$.
The \emph{group} operator $\trueHl{\group_{\Gamma,\Gamma'}}$ groups trees
according to a grouping condition $\Gamma$ and aggregates values of interest
according to $\Gamma'$. Both $\Gamma$ and $\Gamma'$ are sequences of elements
of the form $p \rangle p'$ where $p$ is a path in the input trees, and $p'$ a
path in the output trees.
The \emph{lookup} operator $\trueHl{\lookup_{q=a.r \rangle p}}$ joins input
trees with trees in an external array $a$. The trees to be joined are found by
matching those input trees whose array found applying path $q$ equals the ones
found applying path $r$ to the trees of the external array $a$. The matching
trees from $a$ are stored in the matching input trees under path $p$.

\begin{figure}[tb]
\[
\begin{array}l
\hspace{-5pt}\begin{array}{lll}
s & \gram & 
\hl{\match_\varphi}               
\Div \hl{\unwind_p}               
\Div \hl{\project_\Pi}            
\Div \hl{\group_{\Gamma:\Gamma'}} 
\Div \hl{\lookup_{q=a.r \rangle p}}
\\[.5em]
\varphi & \gram & 
\hl{\true} 
\Div \hl{p = a}
\Div \hl{p_1 = p_2}
\Div \hl{\exists p}
\Div \hl{\neg \varphi}
\Div \hl{\varphi \wedge \varphi}
\Div \hl{\varphi \vee \varphi} 
\\[.5em]
\Pi & \gram & 
\hl{p} 
\Div \hl{d\rangle p}
\Div \hl{p,\Pi}
\Div \hl{d\rangle p,\Pi}
\\[.5em]
d & \gram & 
\hl{b} 
\Div \hl{p}
\Div \hl{[d_1, \cdots, d_n]}
\Div \hl{\varphi}
\Div \hl{\varphi? d_1 : d_2}
\\[.5em]
\Gamma & \gram & \hl{p\rangle p'} \Div \hl{p\rangle p',\Gamma}
\end{array}
\end{array}
\]
\caption{Syntax of the \framework}
\label{fig:syntax}
\end{figure} 
\subsection{Match}
When applied to an array $a$, match $\match_{\varphi}$ returns those elements
in $a$ that satisfy $\varphi$. If there is no element in $a$ that satisfies
$\varphi$, $\match_{\varphi}$ returns an array with no elements (different
from $\emptyarray$). Below, we mark $t \sat \varphi$ the satisfiability of
criterion $\varphi$ by a tree $t$.
\[
\begin{array}{cc}
\emptyarray \app{\match_\varphi} = [\ ]
\hspace{5em}
\begin{array}l  
[t] :: a \app{\match_\varphi} = \begin{cases}
[t] :: (a\app{\match_\varphi}) & \mbox{if } t \sat\varphi
\\
a\app{\match_\varphi} & \mbox{if } \size a > 0
\\
[\ ] & \mbox{otherwise}
\end{cases}
\end{array}
\\[1em]
t \sat \varphi \mbox{ holds iff } \begin{cases}
\varphi = \true
\\
\varphi = (\exists p) \ \wedge \ \apt t p \neq \emptyarray
\\
\varphi = (p = a) \ \wedge\ \apt t p = a
\\
\varphi = (p_1 = p_2) \ \wedge \ t \sat \big( (p_1 = a) \wedge (p_2 = a) \big) 
\\
\varphi = (\neg \varphi') \ \wedge\ t \not\sat\varphi'
\\
\varphi = (\varphi_1 \wedge \varphi_2) \ \wedge \ (t \sat \varphi_1
\ \wedge \ t \sat \varphi_2)
\\
\varphi = (\varphi_1 \vee \varphi_2) \ \wedge \ (t \sat \varphi_1
\ \vee \ t \sat \varphi_2)
\end{cases}
\end{array}
\]
Above, criterion $\varphi = (p_1 = p_2)$ is satisfied both when the
application of the two paths to the input tree $t$ return the same array $a$
as well as when both paths do not exist in $t$, i.e., their application
coincide on $\emptyarray$.

\begin{example}
\label{example:match}
We report below the execution of the \lstinline{match} operator at line 4 of
\Cref{lst:usecase}. In the example, array $a$ corresponds to the data
structure defined at lines 1--2 in \Cref{ex:data_structure}. First we
formalise in \framework the \lstinline{match} operator at line 4: $a
\app{\match_{\varphi}}$ where $$\varphi
=
\lstmath{date} == \lstmath{[20181128\{\}]} \ \vee \ 
\lstmath{date} == \lstmath{[20181129\{\}]} \ \vee \ 
\lstmath{date} == \lstmath{[20181130\{\}]}$$
The match evaluates all trees inside $a$, below we just show that
evaluation for $a[1]$
\begin{lstlisting}[basicstyle=\ttfamily\footnotesize]
a[1] = $\emptyval\ ${date:[20181129{}],t:[37{},...],hr:[ 64 {},...]}  
\end{lstlisting}
we verify if one of the sub-conditions $a[1] \sat \lstmath{date} == 
\lstmath{[20181128\{\}]}$, $a[1] \sat \lstmath{date} == 
\lstmath{[20181129\{\}]}$, or $a[1] \sat \lstmath{date} == 
\lstmath{[20181130\{\}]}$ hold. Each condition is evaluated by applying path
\lstinline{date} on $a[1]$ and by verifying if the equality with the
considered array, e.g., \lstinline|[20181128{}]|, holds.
As a result, we obtain the input array $a$ filtered from the trees that do not
correspond to the dates in the criterion.
\end{example}  
\subsection{Unwind}
To define the semantics of the unwind operator $\unwind$, we introduce the
\emph{unwind expansion operator} $\mult{t}{a}{k}$ (read ``unwind $t$ on $a$
under $k$''). Informally $\mult{t}{a}{k}$ returns an array of trees with
cardinality $\size a$ where each element has the shape of $t$ except that
label $k$ is associated index-wise with the corresponding element in $a$.
Formally, given a tree $t$, an array $a$, and a key $k$:

\vspace{.5em}
\noindent\scalemath{.95}{
\mult{t}{a}{k}
=
\begin{cases}
  \left[ b
  \hl{\Big(\big(\{k_i:a_i\}_i \setminus \{k:k(t)\}\big)} 
  \cup
  \hl{\left\{k : [t']\right\}\Big)} 
  \right] 
  :: \hl{\mult{t}{a'}{k}}
  & \mbox{if } \begin{array}l
    a = [t']::a'
    \\
    \wedge \ t = b \{k_i:a_i\}_i
  \end{array}
\\
[\ ] & \mbox{otherwise}
\end{cases}
}
\vspace{.5em}

\noindent Then, the formal definition of $a\app{\unwind_p}$ is
\[
a\ \app{\ \unwind_p}\begin{cases}
\mult{t}{\apt{t}{k.\emptyseq}\app{\unwind_{p'}}}{k}::a'
\app{\unwind_p}
& \mbox{if } p = e.p' \ \wedge \ \eval{e}{k} \ \wedge \ a = [t]::a'
\\
a & \mbox{if } p = \emptyseq
\\
[\,] & \mbox{otherwise}
\end{cases}
\]
We define the unwind operator inductively over both $a$ and $p$. The induction
over $a$ results in the application of the unwind expansion operator $\multOp$
over all elements of $a$. The induction over $p$ splits $p$ in the current key
$k$ and the continuation $p'$. Key $k$ is used to retrieve the array in the
current element of $a$, i.e., $\apt{t}{k.\emptyseq}{}$, on which we apply
$\unwind_p'$ to continue the unwind application until we reach the
termination with $p = \emptyseq$.

\begin{example}
\label{example:unwind}
We report the execution of the \lstinline{unwind} operator at line 11 of
\Cref{lst:usecase}.

The unwind operator unfolds the given input array wrt a given path $p$ in two
directions. The first is breadth, where we apply the unwind expansion operator
$\mult{t}{a'}{k}$, over all input trees $t$ and wrt the first node $k$ in the
path $p$. The second direction is depth, and defines the content of array $a'$
in $\mult{t}{a'}{k}$, which is found by recursively applying the unwind
operator wrt to the remaining path nodes in $p$ ($k$ excluded) over the arrays
pointed by node $k$ in each $t$.

Let $a$ be the sleep-logs data-structure at lines 4--6 of \Cref
{ex:data_structure}, such that $a = [t_{\text{2018}},t_{\text{2017}},...]$
where e.g., $t_{\text{2018}}$ is that $t$ in $a$ such that
$\apt{t}{\mathtt y}=\lstmath{[2018\{\}]}$. The concatenation below
is the first level of depth unfolding, i.e., for node \lstinline{M} of
unwind $\unwind_{\mathtt{M.D.L}}$.

\mathexample{$
\mult{ t_{\text{2018}} }{ 
  \apt{ t_{\text{2018}} }{ \lstmath{M.}\emptyseq }
  \app{ \unwind_{ \mathtt{D.L} } }
}
{ \lstmath{M} } ::
\mult{ t_{\text{2017}} }{ 
  \apt{ t_{\text{2017}} }{ \lstmath{M.}\emptyseq }
  \app{ \unwind_{ \mathtt{D.L} } }
}
{ \lstmath{M} } :: ...$}

To conclude this example, we show the execution of the unwind expansion
operator $\mult{t_{\text{30}}}{\apt{t_{\text{30}}}{\mathtt{L.}\emptyseq}}{
\mathtt{L}}$ of the terminal node \lstinline{L} in path $p$, relative to the
sleep logs recorded within a day, represented by tree $t_{\text{30}}$, i.e.,
where $\apt{\mathtt{d}}{t_{\text{30}}}=\lstmath{[30\{\}]}$.

\mathexample{\scalemath{.83}{\begin{array}l
\mult{t_{\text{30}}}{\apt{t_{\text{30}}}{\mathtt{L.}\emptyseq}}{\mathtt{L}}
\Rightarrow
\\[0em]
[\emptyval\ (( \{ \mathtt{d}:\lstmath{[30\{\}]}, \mathtt{L}:[\lstmath{...}]\}
\setminus\{\mathtt{L}:[\lstmath{...}]\}) \cup 
\{\lstmath{L}:\lstmath{[} \emptyval \
  \lstmath{\{s:["21:01"\{\}],e:["22:12"\{\}],q:["good"\{\}]\}} \lstmath{]}\})]::
\\[0em] %
[\emptyval\ (( \{ \mathtt{d}:\lstmath{[30\{\}]}, \mathtt{L}:[\lstmath{...}]\}
\setminus\{\mathtt{L}:[\lstmath{...}]\}) \cup 
\{\lstmath{L}:\lstmath{[} \emptyval \
  \lstmath{\{s:["22:36"\{\}],e:["22:58"\{\}],q:["good"\{\}]\}} \lstmath{]}\})]::...
\end{array}
}}

Above, for each element of the array pointed by \lstinline{L}, e.g.,
\lstinline|{s:["21:01"{}],e:["22:12"{}],q:["good"{}]}| we create a
new structure $[\emptyval\ (( \{ \mathtt{d}:\lstmath{[30\{\}]},
\mathtt{L}:[...]\}]$ where we replace the original array associated with the
key \lstinline{L} with a new array containing only that element.
The final result of the \lstinline{unwind} operator has the shape:
\begin{lstlisting}[basicstyle=\ttfamily\footnotesize]
[$\emptyval$ {y:[2018{}],M:[$\emptyval${m:[11{}],D:[$\emptyval$ {d:[30{}],
      L:[$\emptyval${s:"21:01",e:"22:12",q:"good"}]}]}]},
 $\emptyval$ {y:[2018{}],M:[$\emptyval${m:[11{}],D:[$\emptyval$ {d:[30{}],
      L:[$\emptyval${s:"22:36",e:"22:58",q:"good"}]}]}]},... ]
\end{lstlisting}
\vspace{-1em}
\end{example} 
\subsection{Project}
We start by defining some auxiliary operators used in the definition of the
project. Auxiliary operators $\project_p(a)$ and $\project_p(t)$ formalise the
application of a branch-selection over a path $p$. Then, the auxiliary
operator $\evalDef(d,t)$ returns the array resulting from the evaluation of a
definition $d$ over a tree $t$. Finally, we define the projection of a value
(definition) $d$ into a path $p$ over a tree $t$ i.e., $\project_{d \rangle
p}(t)$.
The projection for a path $p$ over an array $a$ results in an array where we
project $p$ over all the elements (trees) of $a$.
\[\project_p([t_1,\cdots,t_n]) = [\ \project_p(t_1),\cdots,\project_p(t_n)\ ]\]
The projection for a path $p$ over a tree $t$ implements the actual semantics
of branch-selection, where, given a path $e.p'$, $\eval{e}{k}$, we remove all
the branches $k_i$ in $t = b\{k_i:a_i\}$, keeping only $k$ (if $k
\in \{k_i\}$) and continue to apply the projection for the continuation $p'$
over the (array of) sub-trees under $k$ in $t$ (i.e., $
\apt{t}{k.\emptyseq}$).
\[
\scalemath{.95}{
\project_p(t) = \begin{cases}
  \emptyval \ \{\ k : \project_{p'}(\apt{t}{k.\emptyseq}) \ \}
  & \mbox{if }
  \apt{t}{p} \neq \emptyarray \ \wedge \
  p = e.p' \ \wedge \ t = b\ \{ k_i : a_i \}_i 
  \ \wedge \ \eval{e}{k}
  \\
  t & \mbox{if } p = \emptyseq
  \\
  \emptytree & \mbox{otherwise}
\end{cases}}
\]
The operator $\evalDef(d,t)$ evaluates the value definition $d$ over the tree
$t$ and returns an array containing the result of the evaluation.
\[
\evalDef(d,t) = \begin{cases}
  [\ d\ \{\,\}\ ] & \mbox{if } d \in V
  \\
  [\ t\sat\varphi\ \{\,\}\ ] & \mbox{if } d \in \varphi
  \\
  \apt{t}{d} & \mbox{if } d \in P
  \\
  \evalDef(d,t)::\evalDef(d',t) & \mbox{if } d = [d]::d'
  \\
  \evalDef(d',t) & \mbox{if } d = \varphi?d_{\true}:d_
  {\false} \ \wedge \ d'=d_{t \sat \varphi}
  \\
  \emptyarray & \mbox{otherwise}
\end{cases}
\]
Then, the application of the projection of a value definition $d$ on a path
$p$, i.e., $\project_{d \rangle p}(t)$ returns a tree where under path $p$ is
inserted the $\evalDef$uation of $d$ over $t$.
\[
\project_{d\rangle p}(t) = \begin{cases}
  \emptyval \ \big\{ k : [\ \project_{d \rangle p'}(t)\ ] \big\} & 
  \mbox{if } p
  = e.p' \ \wedge \eval{e}{k} \ \wedge \ \evalDef(d,t)\neq
  \emptyarray
  \\
  \emptyval \ \big\{ k : \evalDef(d,t) \big\} & \mbox{if } p = e.\emptyseq \ \
  \wedge \eval{e}{k} \ \wedge \ \evalDef(d,t) \neq \emptyarray
  \\
  \emptytree & \mbox{otherwise}
\end{cases}
\]
Before formalising the projection, we define the auxiliary tree-merge operator
$t \merge t'$, used to merge the result of a sequence of projections $\Pi$.
\[
\begin{array}l
([t]::a) \merge ([t']::a') = [t \merge t' ] :: a \merge a'
\qquad
a \merge [\,] = [\,] \merge a = a \merge \emptyarray = \emptyarray \merge a = a
\\[1em]
b\ \{k_i:a_i\}_i \merge b'\ \{k_j:a_j\}_j = 
  \emptytree \quad \mbox{if } b \neq b'
\qquad
t \merge \emptytree = t
\\[1em]
t \merge t' = b \ \{\ k_h : k_h(t) \merge k_h(t') \ \}_{\,h \in I \cup J} 
\quad \mbox{if }
t = b \ \{\ k_i:a_i \ \}_{i \in I} \ \wedge \ 
t' = b \ \{\ k_j:a_j \ \}_{j \in J}
\end{array}
\]
To conclude, first we define the application of the projection to a tree $t$,
i.e., $t\app\project_\Pi$, which merges ($\merge$) into a single tree the
result of the applications of projections $\Pi$ over $t$
\[
\project_\Pi(t) = \begin{cases}
  \project_{p}(t) \merge (t\app\project_{\Pi'}) &
  \mbox{if } \Pi = p,\Pi'
  \\
  \project_{d\rangle p}(t) \merge (t\app\project_{\Pi'}) &
  \mbox{if } \Pi = d\rangle p,\Pi'
  \\
  \project_{p}(t) &
  \mbox{if } \Pi = p
  \\
  \project_{d \rangle p}(t) &
  \mbox{if } \Pi = d \rangle p
\end{cases}
\]
and finally, we define the application of the projection $\project_\Pi$
to an array $a$, i.e., $a\app\project_\Pi$, which corresponds to the
application of the projection to all the elements of $a$.
\[
[\ t_1,\cdots,t_n\ ] \app \project_\Pi = 
  [\ \project_\Pi(t_1),\ \cdots\ ,\project_\Pi(t_n)\ ]
\]

\begin{example}
\label{example:project_1}

We report the execution of the \lstinline{project} at line 5 of
\Cref{lst:usecase}. Let $a$ be the array at the end of \Cref{example:match},
and let $t_{\text{28}}$, $t_{\text{29}}$, $t_{\text{30}}$ be
the trees in $a$ such that $t_{\text{28}}$ is the first tree in
$a$ relative to \texttt{date} \lstinline{20181128}, $t_{\text{29}}$
the second, and $t_{\text{30}}$ the third
\begin{lstlisting}[basicstyle=\ttfamily\footnotesize]
[$t_{\text{28}}$,$t_{\text{29}}$,$t_{\text{29}}$] $
\app{\ \project_{\mathtt{t}\rangle\mathtt{temperatures},\ 
\mathtt{pseudoID}\rangle
\mathtt{patient\_id}}} \Rightarrow$
[$\scalemath{.84}{\project_{\mathtt{t}\rangle\mathtt{temperatures},\ 
\mathtt{pseudoID}\rangle
\mathtt{patient\_id}}(t_{\text{28}})}$,$\scalemath{.84}{\project_{\mathtt{t}\rangle\mathtt{temperatures},\ \mathtt{pseudoID}\rangle
\mathtt{patient\_id}}(t_{\text{29}})}$,$\scalemath{.84}{\project_{\mathtt{t}\rangle\mathtt{temperatures},\ \mathtt{pseudoID}\rangle
\mathtt{patient\_id}}(t_{\text{30}})}$]
\end{lstlisting}
We continue showing the projection of the first element in $a$, $t_{
\text{28}}$ (the projection on the other elements follows the
same structure)
\begin{lstlisting}[basicstyle=\ttfamily\footnotesize]
$\project_{\mathtt{t}\rangle\mathtt{temperatures},\ \mathtt{pseudoID}\rangle
\mathtt{patient\_id}}(t_{\text{28}}) \Rightarrow
\project_{\mathtt{t}\rangle\mathtt{temperatures}}(t_{\text{28}}) \merge
\project_{\mathtt{pseudoID}\rangle
\mathtt{patient\_id}}(t_{\text{28}})\Rightarrow$ 
$\emptyval$ { temperatures : $\project_{\texttt{t}\rangle\emptyseq}(t_{
\text{28}})$ } $\merge$ $\emptyval$ { patient_id : $\project_{
\texttt{["xxx"\{\}]}\rangle\emptyseq}(t_{\text{28}})$ }
$ =\emptyval$ { temperatures : $\evalDef(\texttt{t},t_{\text{28}})$ }$\merge$$
\emptyval$ { patient_id : $\evalDef($["xxx"{}]$,t_{\text{28}})$ }
$ =\emptyval$ { temperatures : $\apt{t_{\text{28}}}{\texttt{t}}$ }$\merge
${patient_id: ["xxx"{}]}
$= \emptyval$ { temperatures : [36{},...], patient_id: ["xxx"{}] }
\end{lstlisting}
The result of the projection has the shape
\begin{lstlisting}[basicstyle=\ttfamily\footnotesize]
[$\emptyval$ { temperatures:[36{},...], patient_id:["xxx"{}] },
 $\emptyval$ { temperatures:[37{},...], patient_id:["xxx"{}] },
 $\emptyval$ { temperatures:[36{},...], patient_id:["xxx"{}] ]
\end{lstlisting}
\end{example} 
\begin{example}
\label{example:project2}
We report the execution of the \lstinline{project} at line 12 of
\Cref{lst:usecase}. Let $a$ be the array at the end of \Cref{example:unwind},
and let $t_{\text{2018}}^{\text{1}}$, $t_{\text{2018}}^{\text{2}}, ...$ be
the trees in $a$ such that $t_{\text{2018}}^{\text{1}}$ is the first tree in
$a$ relative to \texttt{year} \lstinline{2018}, $t_{\text{2018}}^{\text{2}}$
the second, and so on.
\begin{lstlisting}[basicstyle=\ttfamily\footnotesize]
[$t_{\text{2018}}^{\text{1}}$,$t_{\text{2018}}^{\text{2}}$,...] $
\app{\ \project_{\mathtt{y}\rangle\mathtt{year},\ \mathtt{M.m}\rangle
\mathtt{month},\ \mathtt{M.D.d}\rangle\mathtt{day},\ \mathtt{M.D.L.q}\rangle
\mathtt{quality}}} \Rightarrow$
[$\scalemath{.90}{\project_{\mathtt{y}\rangle\mathtt{year},\ 
\mathtt{M.m}\rangle
\mathtt{month},\ \mathtt{M.D.d}\rangle\mathtt{day},\ \mathtt{M.D.L.q}\rangle
\mathtt{quality}}(t_{\text{2018}}^{\text{1}})}$,$\scalemath{.90}{\project_{
\mathtt{y}\rangle \mathtt{year},\ \mathtt{M.m}\rangle \mathtt{month},\ 
\mathtt{M.D.d}\rangle\mathtt{day},\ \mathtt{M.D.L.q}\rangle
\mathtt{quality}}(t_{\text{2018}}^{\text{2}})}$,...]
\end{lstlisting}
We continue showing the projection of the first element in $a$, $t_{
\text{2018}}^{\text{1}}$ (the projection on the other elements follows the
same structure)
\begin{lstlisting}[basicstyle=\ttfamily\footnotesize]
$\project_{\mathtt{y}\rangle\mathtt{year},\ 
\mathtt{M.m}\rangle
\mathtt{month},\ \mathtt{M.D.d}\rangle\mathtt{day},\ \mathtt{M.D.L.q}\rangle
\mathtt{quality}}(t_{\text{2018}}^{\text{1}}) \Rightarrow$
$\project_{\mathtt{y}\rangle\mathtt{year}}(t_{\text{2018}}^{\text{1}}) \merge
\project_{\mathtt{M.m}\rangle\mathtt{month}}(t_{\text{2018}}^{\text{1}}) \merge
\project_{\mathtt{M.D.d}\rangle\mathtt{day}}(t_{\text{2018}}^{\text{1}}) \merge
\project_{\mathtt{M.D.L.q}\rangle\mathtt{quality}}(t_{\text{2018}}^{\text{1}})$
\end{lstlisting}
Finally, we show the unfolding of the first two projections from the left,
above, i.e., those for $\mathtt{y}\rangle\mathtt{year}$ and for $
\mathtt{M.m}\rangle\mathtt{month}$, and their merge $\merge$ (the
remaining ones unfold similarly).
\begin{lstlisting}[basicstyle=\ttfamily\footnotesize]
$\project_{\mathtt{y}\rangle\mathtt{year}}(t_{\text{2018}}^{\text{1}}) \merge
\project_{\mathtt{M.m}\rangle\mathtt{month}}(t_{\text{2018}}^{\text{1}}) \Rightarrow$
$\emptyval$ { year : $\project_{\texttt{y}\rangle\emptyseq}(t_{\text{2018}}^{
\text{1}})$ } $\ \merge$ $\emptyval$ { month : $\ \project_{
\texttt{M.m}\rangle\emptyseq}(t_{\text{2018}}^{\text{1}})$ }
= $\emptyval$ { year : $\evalDef(\texttt{y},t_{\text{2018}}^{\text{1}})$ } $\
\merge$ $
\emptyval$ { month : $\ \evalDef(\texttt{M.m},t_{\text{2018}}^{\text{1}})$ }
= $\emptyval$ { year : $\apt{t_{\text{2018}}^{\text{1}}}{
\texttt{y}}$ }$\merge$$
\ \emptyval$ { month : $\apt{t_{\text{2018}}^{\text{1}}}{\texttt{M.m}}$ }
= $\emptyval$ { year : [2018{}] } $\merge$ $\emptyval$ { month : [11{}] } 
= $\emptyval$ { year : [2018{}], month : [11{}] }
\end{lstlisting}
The result of the projection has the shape
\begin{lstlisting}[basicstyle=\ttfamily\footnotesize]
[$\emptyval${year:[2018{}],month:[11{}],day:[30{}],quality:["good"{}]},
 $\emptyval${year:[2018{}],month:[11{}],day:[30{}],quality:["good"{}]},
 $\emptyval${year:[2018{}],month:[11{}],day:[30{}],quality:["poor"{}]},...]
\end{lstlisting}
\end{example}

\subsection{Group}
The group operator takes as parameters two sequences of paths, separated by
a semicolon, i.e., $q_1 \rangle p_1,\cdots,q_n \rangle p_n : s_1\rangle
r_1,\cdots,s_m\rangle r_m$. The first sequence of paths, ranged $[1,n]$, is
called \emph{aggregation set}, while the second sequence, ranged $[1,m]$,
is called \emph{grouping set}.
Intuitively, the group operator first groups together the trees in $a$
which have the maximal number of paths $s_1,\cdots,s_m$ in the grouping set
whose values coincide. The values in $s_1,\cdots,s_m$ are projected in the
corresponding paths $r_1,\cdots, r_m$. Once the trees are grouped, the
operator aggregates all the different values, without duplicates, found in
paths $q_1,\cdots,q_n$ from the aggregation set, projecting them into the
corresponding paths $p_1,\cdots,p_n$.
We start the definition of the grouping operator by expanding its application
to an array $a$. In the expansion below, on the right, we use the
series-concatenation operator $\bigop{::}$ and the set $H$, element of the
power set $2^{[1,n]}$, to range over all possible combinations of paths in the
grouping set. Namely, the expansion corresponds to the concatenation of all
the arrays resulting from the application of the group operator on a subset
(including the empty and the whole set) of paths in the grouping set.
\[
\group_{q_1 \rangle p_1 ,\cdots,q_n \rangle p_n  :s_1 \rangle r_1 , \cdots,
s_m \rangle r_m } \app a 
=
\bigop{::}\limits_{\forall H \in 2^{[1,m]}}
\group^{H}_{q_1 \rangle p_1 ,\cdots,q_n \rangle p_n  :s_1 \rangle r_1 , \cdots,
s_m \rangle r_m }(a)
\]
In the definition of the expansion, we mark $\castToSet{a}$ the casting of an
array $a$ to a set (i.e., we keep only unique elements in $a$ and lose their
relative order). Each $\group^{H}_{q_1\rangle p_1,\cdots,q_n\rangle p_n
:s_1\rangle r_1, \cdots, s_m\rangle r_m}(a)$ returns an array that contains
those trees in $a$ that correspond to the grouping illustrated above.
Formally:
\[
\begin{array}l  
\group_{
q_1\rangle p_1,\cdots,q_n\rangle p_n :s_1\rangle r_1, \cdots,
s_m\rangle r_m}^{H}(a) =\\[1em]
\begin{cases}
\bigop{::}\limits_{\forall a\!'}
\left[
\bigop{\bigoplus}\limits_{i=1}^{n}
\project_{\chi,\theta_i}
(\emptytree)
\right]
&
\mbox{if } 
\begin{array}l
  h \in H \ \wedge \ a\!'[h] \in \Big\{ 
    \apt{t}{s_h} \ | \ t \in \castToSet{a} \wedge \apt{t}{s_h}
    \neq \emptyarray
  \Big\}
  \\
  \wedge \ \chi = (a\!'[h]\rangle H,[r_1,\cdots,r_n])
  \\ \wedge \ \theta_i = \bigop{::}\limits_{\forall t_i}\apt{t_i}
  {q_i} \rangle p_i
  \
  \wedge \ t_i \in \castToSet{a \app \match_{\psi_i}} \supset \emptyset
  \\
  \wedge \ \psi_i = 
  \exists q_i 
  \ \wedge\ 
  \neg \bigop{\bigvee}\limits_{j \not \in H} \exists s_j
  \ \wedge \bigop{\bigwedge}\limits_{h}
  \Big( \big( s_h = a'[h] \big) \wedge \exists s_h \ \Big)
\end{array}
\\
[\ ] & \mbox{otherwise}
\end{cases}
\end{array}
\]
When applied over a set $H$, $h \in H$, $\group$ considers \emph{all}
combinations of values identified by paths $s_h$ in the trees in $a$. In the
formula above, we use the array $a\!'$ to refer to those combinations of
values. In the definition, we impose that, for each element in $a\!'$ in a
position $h$, there must be at least one tree in $a$ that has a non-null
($\neq \emptyarray$) array under path $s_h$. Hence, for each combination
$a\!'$ of values in $a$, $\group$ builds a tree that \emph{i)} contains under
paths $r_h$ the value $a\!'[h]$ (as encoded in the projection query $\chi$ and
from the definition of the operator $a\!'[h]\rangle H, a $, defined below) and
\emph{ii}) contains under paths $p_i$, $1\leq i \leq n$, the array containing
all the values found under the correspondent path $q_i$ in all trees in $a$
that match the same combination element-path in $a\!'$ (as encoded in
$\theta_i$). The grouping is valid (as encoded in $\psi_i$) only if we can
find (i.e., match $\match$) trees in $a$ where \emph{i}) we have a non-empty
value for $q_i$, \emph{ii}) there are no paths $s_j$ that are excluded in $H$,
and \emph{iii}) for all paths considered in $H$, the value found under path
$s_h$ corresponds to the value in the considered combination $a'[h]$. If the
previous conditions are not met, $\group$ returns an empty array $[\ ]$.

We conclude defining the operator $a\!'[h] \rangle H, a$, used above to unfold
the set of aggregation paths and the related values contained in $H$, e.g.,
let $H = \{1,3,5\}$ then $a\!'[h] \rangle H, a  =  a'[1]\rangle p_1,a'
[3]\rangle p_3 ,a'[5]\rangle p_5$. Its meaning is that, for each path
$p_\bullet$, we project in it the value correspondent to $a'[\bullet]$.
Formally
\[
a\!'[h]\rangle H, a  = \begin{cases}
   a\!'[j]\rangle a[j], (a\!'[h] \rangle (H\setminus \{j\}), a) 
    & \mbox{if } |H| > 1 \wedge j \in H
  \\
  a\!'[j] \rangle a[j] & \mbox{if } |H| = 1 \wedge j \in H
  \\
  \emptyseq & \mbox{otherwise}
\end{cases}
\]

Note that for case $\group^{\emptyset}_{\cdots}$ (i.e., for $H =
\emptyset$), $a\!'[h]\rangle H, a$ returns the empty path $\emptyseq$,
which has no effect (i.e., it projects the input tree) in the projection
$\project_\chi$ in the definition of $\group$. Hence, the resulting tree from
grouping over $\emptyset$ will just include (and project over
$p_1,\cdots,p_n$) those trees in $a$ that do not include any value reachable
by paths $s_1,\cdots,s_m$ (as indicated by expression $\neg
\bigop{\bigvee}\limits_{j \not \in H} \exists s_j$ in $\psi_i$).

Like in MQuery and MongoDB, we allow the omission of paths
$p_1,\cdots,p_n$ and $r_1,\cdots,r_n$ in $\Gamma:\Gamma'$. However, we
interpret this omission differently wrt MQuery. There,
the values obtained from $q_i$s with missing $p_i$s (resp., $s_i$ with missing
$r_i$) are stored within a default path \lstinline{_id}. Here, we intend the
omission as an indication of the fact that the user wants to preserve the
structure of $q_i$ (resp., $s_i$) captured by the structural equivalence
below.
\[
\group_{q_1,\cdots,q_n:s_1,\cdots,s_m} \equiv
\group_{q_1\rangle q_1,\cdots,q_n\rangle q_n:s_1\rangle s_1,\cdots,s_m\rangle s_m} 
\]

\begin{example}
We report the execution of the \lstinline{group} operator at line 14 of
\Cref{lst:usecase}. Let $a$ be the result of the projection \Cref
{example:project2}, with the exception that $a$ has been filtered by the
\lstinline{match} at line 13 in \Cref{lst:usecase} and contains only the
sleep logs for \lstinline{day}s \lstinline{29} and \lstinline{30} of 
\lstinline{month} \lstinline{11} and \lstinline{year} \lstinline{2018}.

\mathexample{\vspace{-1em}\[\begin{array}l
a \app{\group_{\texttt{quality} \ :\ \texttt{day},\texttt{month},
\texttt{year}}} \equiv a \app{\group_{\texttt{quality}\rangle\texttt{quality} \ :\ 
\texttt{day}\rangle\texttt{day},\texttt{month}\rangle\texttt{month},
\texttt{year}\rangle\texttt{year}}}
\\[.5em]
\Rightarrow \bigop{::}\limits_{\forall H \in 2^{[1,1]}}
\group^{H}_{\texttt{quality}\rangle\texttt{quality}\ :\
\texttt{day}\rangle\texttt{day},\texttt{month}\rangle\texttt{month},
\texttt{year}\rangle\texttt{year}}(a) 
\\[1.2em]
= \left[\group^{\emptyset}_{\texttt{quality}\rangle\texttt{quality}\ :\
\texttt{day}\rangle\texttt{day},\texttt{month}\rangle\texttt{month},
\texttt{year}\rangle\texttt{year}}(a)\right]
\quad\begin{array}l
\mbox{\small\color{darkgray}this to equals $[\ ]$}\\
\mbox{\small\color{darkgray}since $\psi_i$ is always false in $a$}
\end{array}
\\\;\quad::\left[\group^{\{1\}}_{\texttt{quality}\rangle\texttt{quality}\ :\
\texttt{day}\rangle\texttt{day},\texttt{month}\rangle\texttt{month},
\texttt{year}\rangle\texttt{year}}(a)\right]
\\[.6em] 
= [\ ] :: \left[ \project_{\scalemath{.7}{\lstmath{[30
\{\}]}\rangle\lstmath{day},\lstmath{[11
\{\}]}\rangle\lstmath{month},\lstmath{[2018
\{\}]}\rangle\lstmath{year} }}(\emptytree)
\merge
\project_{\scalemath{.7}{\lstmath{quality}\rangle \lstmath{["good"\{\},"good"
\{\},...]}}}(\emptytree)
\right]
\\[.5em] \hspace{2.15em} 
:: \left[ \project_{\scalemath{.7}{\lstmath{[29
\{\}]}\rangle\lstmath{day},\lstmath{[11
\{\}]}\rangle\lstmath{month},\lstmath{[2018
\{\}]}\rangle\lstmath{year} }}(\emptytree)
\merge
\project_{\scalemath{.7}{\lstmath{quality}\rangle \lstmath{["good"\{\},"poor"
\{\},...]}}}(\emptytree)
\right]
\\[.5em]
= \scalemath{.88}{\lstmath{[}\emptyval\ 
\lstmath{\{ day: [30\{\}],month: [11\{\}],year: [2018\{\}]}
\lstmath{,quality: ["good"\{\},"good"\{\},...]\},}}
\\
\hspace{1.5em}\scalemath{.88}{\emptyval\ \lstmath{\{ day: [29\{\}],month: [11
\{\}],year: [2018\{\}]}\lstmath{,quality: ["good"\{\},"poor"\{\},...]\}}
\lstmath{]}}
\end{array}
\]}

\end{example} 
\subsection{Lookup}
Informally, the lookup operator joins two arrays, a source $a$ and an adjunct
$a'$, wrt a destination path $p$ and two source paths $q$ and $r$. Result of
the lookup is a new array that has the shape of the source array $a$ but where
each of its elements $t$ has under path $p$ those elements in the adjunct
array $a'$ whose values under path $r$ equal the values found in $t$ under
path $q$. Formally
\[\begin{array}l
a\ \app\ \lookup_{q=a'.r\rangle p} = 
[\ \project_{\emptyseq,\beta_{1}}(a[1])\ ]
:: \cdots :: 
[\ \project_{\emptyseq,\beta_{n}}(a[n])\ ]
\mbox{ s.t. }
\begin{cases}
\beta_i = ( a' \app \match_{r = a''} ) \rangle p \\
\wedge \ a'' = \apt{a[i]}{q} \\
\wedge \ 1 \leq i \leq n
\end{cases}
\end{array}\]
Above, the lookup operator $\lookup$ takes as parameters three
paths $p$, $q$, and $r$ and an array of trees $a'$. When applied to an array
of trees $a = [t_1,\cdots,t_n]$, it returns $a$ (i.e., all of its elements, as
retuned by the projection $\project$ under the first parameter $\emptyseq$)
where each of its elements has under path $p$ an array of trees obtained from
applying the match ($\match_{r = a''}$) in expression $\beta_i$, i.e.,
following the definition of $\project$, the projection under $\emptyseq$ is
merged with the result of the projection under $\beta_i$. For each element
$a[i]$ ($1\leq i \leq n$), $\beta_i$ matches those trees in $a'$ for which
either \emph{i}) there is a path $r$ and the array reached under $r$ equals
the array found under $\apt{a[i]}{q}$ or \emph{ii}) there exist no path $r$
(i.e., its application returns the null array $\emptyarray$) and also $q$ does
not exist in $t_i$ (i.e., $\apt{a[i]}{q} = \emptyarray$).
\begin{example}
We report the execution of the \lstinline{lookup} at line 16 of \Cref{lst:usecase}
$$\begin{array}l
a \app \lookup_{ \mathtt{patient\_id} = a'\mathtt{.patient\_id} \rangle
\mathtt{temps} }
\end{array}$$
where $a$ corresponds to the resulting array from the application of
the \lstinline{project} operator at line 15 of \Cref{lst:usecase}, which has
the shape
\begin{lstlisting}[basicstyle=\ttfamily\footnotesize]
$a$ = [  $\emptyval$ { quality:["good"{},"good"{},...], patient_id:["xxx"{}] },
      $\emptyval$ { quality:["poor"{},"good"{},...], patient_id:["xxx"{}] } ]
\end{lstlisting}
and where $a'$ corresponds to the array of temperatures that results from the
application of the \lstinline{project} at line 5 of \Cref{lst:usecase}, as
shown at the bottom of \Cref{example:project_1}.
Then, unfolding the execution of the \lstinline{lookup}, we obtain the
concatenation of the results of two projections, on the only two elements in
$a$. The first corresponds to the projection on $\emptyseq,\beta_1$ while the
second corresponds to the projection on $\emptyseq,\beta_2$ where
$$\begin{array}l 
\beta_1 = \beta_2 = a' \app \match_{\mathtt{patient\_id} = \texttt{["xxx"
\{\}]}} \rangle
\texttt{temps}
\end{array}$$
Below, sub-node \texttt{temps} contains the whole array $a'$,
since all its elements match \texttt{patient\_id}.
\begin{lstlisting}[basicstyle=\ttfamily\footnotesize]
[$\project_{\emptyseq,\beta_1}(a[1])$::$\project_{\emptyseq,\beta_2}(a[2])$]
= [$\emptyval$ { quality:["good"{},"good"{},...], patient_id:["xxx"{}],
    temps: [
       $\emptyval$ { temperatures:[36{},...], patient_id:["xxx"{}] },
       $\emptyval$ { temperatures:[37{},...], patient_id:["xxx"{}] },
       $\emptyval$ { temperatures:[36{},...], patient_id:["xxx"{}] } ] },
 $\emptyval$ { quality:["poor"{},"good"{},...], patient_id:["xxx"{}],
    temps: [
       $\emptyval$ { temperatures:[36{},...], patient_id:["xxx"{}] },
       $\emptyval$ { temperatures:[37{},...], patient_id:["xxx"{}] },
       $\emptyval$ { temperatures:[36{},...], patient_id:["xxx"{}] } ] } ]
\end{lstlisting}
\end{example}  
\section{Related Work and Conclusion}
\label{sec:conclusion}
In this paper, we focus on ephemeral data handling and contrast DBMS-based
solutions wrt to integrated query engines within a given application memory.
We indicate issues that make unfit DBMS-based solutions in ephemeral
data-handling scenarios and propose a formal model, called \framework, to
express document-based queries over common (JSON, XML, ...), tree-shaped data
structures.

\framework instantiates MQuery~\cite{botoeva18}, a sound variant of the
Aggregation Framework~\cite{agg_framework} used in MongoDB, one of the main
NoSQL DBMSes for document-oriented queries.

We implemented \framework in Jolie, a language to program native
microservices, the building blocks of modern systems where ephemeral data
handling scenarios are becoming more and more common, like in
Internet-of-Things, eHealth, and Edge Computing architectures. Jolie offers
variety-by-construction, i.e., the language runtime automatically and
efficiently handles data conversion, and all Jolie variables are trees. These
factors allowed us to separate input/output data-formats from the
data-handling logic, hence providing programmers with a single, consistent
interface to use \framework on any data-format supported by Jolie.

In our treatment, we presented a non-trivial use case from eHealth, which
provide a concrete evaluation of both \framework and MQuery, while also
serving as a running example to illustrate the behaviour of the \framework
operators.

Regarding related work, we focus on NoSQL systems, which either target
documents, key/value, and graphs. The NoSQL systems closest to ours are the
MongoDB~\cite{mongodbwebsite} Aggregation Framework, and the
CouchDB~\cite{couchdbwebsite} query language which handle JSON-like documents
using the JavaScript language and REST APIs. ArangoDB~ \cite{arangodbwebsite}
is a native multi-model engine for nested structures that come with its own
query language, namely ArangoDB Query Language. Redis~\cite{redis} is an
in-memory multi-data-structure store system, that supports string, hashes,
lists, and sets, however it lacks support for tree-shaped data. We conclude
the list of external DB solutions with Google Big
Table~\cite{chang2008bigtable} and Apache HBase~\cite{george2011hbase} that
are NoSQL DB engines used in big data scenarios, addressing scalability
issues, and thus specifically tailored for distributed computing. As argued in
the introduction, all these systems are application-external query execution
engine and therefore unfit for ephemeral data-handling scenarios.

There are solutions that integrate linguistic abstractions to query data
within the memory of an application. One category is represented by
Object-relation Mapping (ORM) frameworks~\cite{fussel1997foundations}.
However, ORMs rely on some DBMS, as they map objects used in the application
to entities in the DBMS for persistence. Similarly,
Opaleye~\cite{ellis2014opaleye} is a Haskell library providing a DSL
generating PostgreSQL\@. Thus, while being integrated within the application
programming tools and executing in-memory, in ephemeral data-handling
scenarios, ORMs are affected by the same issues of DBMS systems. Another
solution is LevelDB~\cite{leveldbwebsite}, which provides both a on-disk and
in-memory storage library for C++, Python, and Javascript, inspired by Big
Table and developed by Google, however it is limited to key-value data
structures and does not support natively tree-shaped data. As cited in the
introduction, a solution close to ours is LINQ~\cite{meijer2006linq}, which
provides query operators targeting both SQL tables and XML nested structures
with \textit{.NET} query operators. Similarly, CQEngine~\cite{cqenginewebsite}
provides a library for querying Java collections with SQL-like operators. Both
solutions do not provide automatic data-format conversion, as our
implementation of \framework in Jolie.

We are currently empirically evaluating the performance of our implementation
of \framework in application scenarios with ephemeral data handling
(Internet-of-Things, eHealth, Edge Computing). The next step would be to use
those scenarios to conduct a study comparing our solution wrt other proposals
among both DBMS and in-memory engines, evaluating their impact on performance
and the development process. Finally, on the one hand, we can support new data
formats in Jolie, which makes them automatically available to our \framework
implementation. On the other hand, expanding the set of available operators in
\framework would allow programmers to express more complex queries over any
data format supported by Jolie. 
\bibliographystyle{ieeetr}
\bibliography{biblio}

\end{document}